\documentclass[pra,showpacs,amsmath,amssymb, twocolumn, 10pt]{revtex4}

\usepackage{amsthm}
\usepackage{dcolumn}
\usepackage{bm}
\usepackage{graphicx}

\begin{document}

\newtheorem{corollary}{Corollary}
\newtheorem{definition}{Definition}
\newtheorem{example}{Example}
\newtheorem{lemma}{Lemma}
\newtheorem{proposition}{Proposition}
\newtheorem{theorem}{Theorem}
\newtheorem{fact}{Fact}
\newtheorem{property}{Property}
\newcommand{\bra}[1]{\langle #1|}
\newcommand{\ket}[1]{|#1\rangle}
\newcommand{\braket}[3]{\langle #1|#2|#3\rangle}
\newcommand{\ip}[2]{\langle #1|#2\rangle}
\newcommand{\op}[2]{|#1\rangle \langle #2|}

\newcommand{\tr}{{\rm tr}}
\newcommand{\supp}{{\rm Supp}}
\newcommand{\sch}{{\rm Sch}}

\newcommand {\E } {{\mathcal{E}}}
\newcommand {\F } {{\mathcal{F}}}
\newcommand {\diag } {{\rm diag}}

\title{Locally Indistinguishable Subspaces Spanned by Three-Qubit Unextendible Product Bases}
\author{Runyao Duan$^{1}$}
\email{dry@tsinghua.edu.cn}
\author{Yu Xin$^{1,2}$}
\email{xiny05@mails.tsinghua.edu.cn}
\author{Mingsheng Ying$^{1}$}
\email{yingmsh@tsinghua.edu.cn}

\affiliation{$^1$State Key Laboratory of Intelligent Technology and
Systems, Tsinghua National Laboratory for Information Science and
Technology, Department of Computer Science and Technology, Tsinghua
University, Beijing 100084, China,
\\
$^2$Department of Physics, Tsinghua University, Beijing 100084,
China}

\date{\today}

\begin{abstract}
We study the local distinguishability of general multi-qubit states
and show that local projective measurements and classical
communication are as powerful as the most general local measurements
and classical communication. Remarkably, this indicates that the
local distinguishability of multi-qubit states can be decided
efficiently. Another useful consequence is that a set of orthogonal
$n$-qubit states is locally distinguishable only if the summation of
their orthogonal Schmidt numbers is less than the total dimension
$2^n$. When $n=2$ such a condition is also sufficient. Employing
these results, we show that any orthonormal basis of a subspace
spanned by arbitrary three-qubit orthogonal unextendible product
bases (UPB) cannot be exactly distinguishable by local operations
and classical communication. This not only reveals another intrinsic
property of three-qubit orthogonal UPB, but also provides a class of
locally indistinguishable subspaces with dimension $4$. We also
explicitly construct locally indistinguishable subspaces with
dimensions $3$ and $5$, respectively. In particular, $3$ is the
minimal possible dimension of locally indistinguishable subspaces.
Combining with the previous results, we conclude that any positive
integer between $3$ and $7$ is the possible dimension of some
three-qubit locally indistinguishable subspace.
\end{abstract}


\maketitle
\section{Introduction}
An interesting problem in quantum information theory is to
distinguish a finite set of orthogonal multipartite quantum states
using local operations and classical communication (LOCC). This
problem has been extensively studied in the last two decades and
numerous exciting results have been reported, see Ref. \cite{DFXY07}
and references therein for details. In spite of these considerable
efforts, a complete solution to this problem is still out of reach
due to the complicated nature of LOCC operations. Nevertheless,
several objects with rich mathematical structure such as
unextendible product bases (UPB) and locally indistinguishable
subspaces were introduced during this process.

The notion of UPB was originally introduced by Bennett and coworkers
\cite{BDM+99}, and has been thoroughly studied in the literatures
\cite{DMS+00, AL01, NC06}. Notably, the members of a UPB cannot be
perfectly distinguishable by LOCC.

Recently Watrous introduced a class of interesting bipartite
subspaces having no orthonormal bases distinguishable by LOCC
\cite{WAT05}. Such subspaces can be intuitively named locally
indistinguishable subspaces. The minimal dimension of locally
indistinguishable subspaces obtained by Watrous is $8$. In Ref.
\cite{DFXY07} we generalized  his result to multipartite setting ans
showed that any orthogonal complement of a multipartite pure state
with orthogonal Schmidt number at least $3$ is locally
indistinguishable. Furthermore, a $3\otimes 3$ subspace of dimension
$7$ and a $2\otimes 2\otimes 2$ subspace of dimension $6$ were
constructed. All these subspaces are actually indistinguishable by a
wider class of quantum operations, say separable operations.
However, it still remains unknown how low can the dimension of
indistinguishable subspaces go. In general, there is no simple and
feasible way to show a given subspace is locally indistinguishable.

In this paper we try to connect the above two notions together. The
problem we attempt to solve can be described as follows. Let $S$ be
an orthogonal UPB, and let $span(S)$ be the subspace spanned by the
members of $S$. Our ultimate goal is to show that $span(S)$ is
locally indistinguishable. Unfortunately, we fail to present a
complete solution at present. We can only partially accomplish this
goal by showing that any subspace spanned by the members of a
$2\otimes 2\otimes 2$ UPB is locally indistinguishable. These
subspaces also have the interesting property that they are
indistinguishable by LOCC but have at least an orthonormal basis
distinguishable by separable operations. As a direct consequence, we
obtain a class of locally indistinguishable subspaces with dimension
$4$.

It should be noted that unextendibility itself is not the crucial
property for local indistinguishability. To see this, we present an
explicit $2\otimes 2\otimes 2$ subspace which is spanned by a set
locally distinguishable quantum states. As an interesting byproduct,
we show the state space of three-qubits can be decomposed into two
orthogonal subspaces which both are entangled, thus are
unextendible. That immediately yields an instance of two orthogonal
mixed states that are locally indistinguishable, even
probabilistically, which is strikingly different from the perfect
local distinguishability of any two orthogonal pure states
\cite{WSHV00}. To the best of our knowledge, no such instance was
previously known.

Since any two orthogonal pure states are perfectly distinguishable
by LOCC \cite{WSHV00}, the dimension of locally indistinguishable
subspace is at least $3$. However, it seems a formidable task to
find such a subspace. Actually for $3\otimes 3$ state space it was
conjectured that such 3-dimensional subspace does not exist at all
\cite{KM05}. Interestingly, for three-qubit system we do find a
class of three-dimensional subspaces that are locally
indistinguishable. This is the first locally indistinguishable
subspace with minimal dimension.

It remains unknown whether there is a locally indistinguishable
subspace with dimension $5$. We provide an affirmative answer to
this question by explicitly constructing a five-dimensional
three-qubit subspace containing a unique product state. Then this
subspace is locally indistinguishable follows from the
Schmidt-number-summation criterion about the local
distinguishability for multi-qubit system.

The rest of this paper is organized as follows. In Section II we
review basic definitions and notions that are useful in studying
LOCC discrimination. Then in Section III a criterion for the local
distinguishability of general $2\otimes n$ orthogonal quantum states
is given, which slightly generalizes the work of Walgate and Hardy
\cite{WH02}. We apply this criterion in Section IV to show that
local projective measurements and classical measurements are
sufficient to perfectly distinguish multi-qubit orthogonal quantum
states. As a direct consequence, we show that a set of multi-qubit
orthogonal quantum states is locally distinguishable only if the
total summation of their orthogonal Schmidt numbers is not more than
the dimension of the state space under consideration. In particular,
in Section V we further prove that the summation of orthogonal
Schmidt numbers less than $4$ gives exactly the necessary and
sufficient condition of the local distinguishability of $2\otimes 2$
orthogonal quantum states. With above preparations, we are ready to
present our main result in Section VI. That is, any subspace spanned
by a three-qubit UPB is locally indistinguishable. Section VII
clarifies the relation between indistinguishablity and
unextendibility. In particular, two $2\otimes 2\otimes 2$ orthogonal
mixed states that are indistinguishable by probabilistic LOCC are
presented. Locally indistinguishable subspaces with dimensions $3$
and $5$ are presented in Sections VIII and IX, respectively.  We
conclude the paper in Section X. Several unsolved problems are also
proposed for further study.
\section{Preliminaries}
We consider a multipartite quantum system consisting of $K$ parts,
say $A_1,\cdots, A_K$. We assume part $A_k$ has a state space
$\mathcal{H}_k$ with dimension $d_k$. The whole state space is given
by $\mathcal{H}=\otimes_{k=1}^K\mathcal{H}_k$ with total dimension
$D=d_1\cdots d_K$. The notation $d_1\otimes \cdots \otimes d_K$ is
an abbreviation for $\mathcal{H}$.

We first recall some useful definitions introduced by Walgate and
Hardy \cite{WH02}. When discriminating a set of states, we need to
perform suitable measurements $\{M_m\}$ on the system to obtain
useful information about the real identity of the states. If
$M^\dagger_mM_m$ is proportional to the identity, then the
measurement operator $M_m$ is simply a unitary operation on the
state and it cannot provide any useful information for
discrimination. This kind of measurement operator is said to be
trivial. A measurement is said to be non-trivial if at least one of
its measurement operators is not trivial.

Any finite LOCC protocol that discriminates a set of  orthogonal
multipartite states $\{\rho_1,\cdots,\rho_N\}$ consists of finitely
many \textit{measuring and broadcasting} rounds as follows: Some
party performs a measurement and then broadcasts the outcome through
the classical channels to the others. The protocol will not
terminate until a definite decision about the identity of the state
can be made. Clearly, there should be some person who performs the
first non-trivial measurement. This simple observation leads us to
the following definition:

\begin{definition}\label{first-measure}\upshape
Suppose Alice, Bob, $\cdots$, try to discriminate a set of states
among them by local operations and classical communication only. We
say Alice goes first if Alice is the one who performs the first
non-trivial measurement.
\end{definition}

After performing a non-trivial measurement, Alice, Bob, $\cdots$,
need to discriminate a new set of (unnormalized) states $\{M_m
\rho_k M_m^\dagger: k=1,\cdots, N\}$ if the outcome is $m$.  To
ensure a perfect discrimination can be achieved, the resulting
states should be orthogonal and some states may be vanishing. This
puts a strong constraint on the measurement operators.
\begin{definition}\label{ortho-keep}\upshape
Let $\{\rho_1,\cdots, \rho_N\}$ be a set of orthogonal states.  A
measurement operator $M_m$ is called orthogonality-keeping if states
in the set $\{M_m\rho_k M_m^\dagger: k=1,\cdots, N\}$ remain to be
orthogonal. A complete measurement $\{M_m\}$ is said to be
orthogonality-keeping if for each $M_m$ is orthogonality-keeping.
\end{definition}

For a positive operator $\rho$, $\supp(\rho)$ represents the support
of $\rho$. In other words, $\supp(\rho)$ is the subspace spanned by
the eigenvectors of $\rho$ corresponding to the positive
eigenvalues. We shall need the following technical lemma:
\begin{lemma}\upshape
Let $\rho$ be a density operator, and $\ket{\psi}$ be a normalized
state. Then $\ket{\psi}\in Supp(\rho)$ if and only if there exists
$0<p\leq 1$ and a density operator $\rho'$ such that
$\rho=p\op{\psi}{\psi}+(1-p)\rho'$.
\end{lemma}

The physical meaning of the above lemma can be interpreted as
follows: A pure state $\ket{\psi}$ is in the support of $\rho$ if
and only if it appears in some ensemble that realizes $\rho$. A
useful consequence is as follows:
\begin{corollary}\label{orthogonality-keeping}\upshape
A measurement $\{M_m\}$ is orthogonality-keeping for a set of
orthogonal states $\{\rho_k\}$ if and only if for each $m$,
$M_m\ket{\psi_1},\cdots, M_m\ket{\psi_N}$ are pairwise orthogonal,
where $\ket{\psi_k}\in \supp(\rho_k)$.
\end{corollary}

In what follows we shall frequently employ the notion of orthogonal
Schmidt number. Here we simply state a definition, for details we
refer to Ref. \cite{DFXY07}. Let $\rho$ be a general quantum state.
The orthogonal Schmidt number of $\rho$, denoted as
$\sch_\perp(\rho)$, is the minimal number orthogonal product states
needed to span the support of $\rho$. When only bipartite pure
states are involved, the orthogonal Schmidt number is exactly the
ordinary Schmidt number, i.e., the number of nonzero Schmidt
coefficients.

\section{A criterion for the Local distinguishability of general orthogonal $2\otimes n$ states}
Now we turn to study the local distinguishability of $2\otimes n$
quantum system. Without loss of generality, we assume Alice is the
person who holds the qubit. In Ref. \cite{WH02}, Walgate and Hardy
gave a necessary and sufficient condition for the local
distinguishability of a finite set of $2\otimes n$ pure states when
Alice goes first. Interestingly, their result is also valid for
mixed states.
\begin{theorem}\label{2otimesn}\upshape
Let Alice and Bob share an unknown state which is secretely chosen
from $N$ orthogonal $2\otimes n$ states, say $\rho_1,\cdots,\rho_N$,
and let Alice be the one holding the qubit. If Alice goes first,
then Alice and Bob can perfectly identify their state using LOCC if
and only if there exists an orthogonal basis $\{\ket{0},\ket{1}\}_A$
such that for any $\ket{\psi_k}\in Supp(\rho_k)$, we have
\begin{equation}\label{nice-form}
\ket{\psi_k}=\ket{0}_A\ket{\psi_0^{(k)}}_B+\ket{1}_A\ket{\psi_1^{(k)}}_B,
\end{equation}
where
$\ip{\psi_0^{(k)}}{\psi_0^{(l)}}=\ip{\psi_1^{(k)}}{\psi_1^{(l)}}=0$
for any $1\leq k<l\leq N$.
\end{theorem}

\textbf{Proof}. Many proof techniques are borrowed from Ref.
\cite{WH02}. The sufficiency is obvious. If there exists an
orthonormal basis $\{\ket{0},\ket{1}\}_A$ such that the above
equation holds, then each $\rho_k$ should be of  the following form:
$$\rho_k=\sum_{m,n=0}^1\op{m}{n}\otimes \rho_{mn}^{(k)},$$
where $\{\rho_{mm}^{(k)}\}$ is a set of orthogonal states and
$m=0,1$. A perfect discrimination protocol is as follows:

(1) Alice measures her qubit according to the basis
$\{\ket{0},\ket{1}\}_A$ and sends the outcome $m$ to Bob.

(2) If $m=0$ then Bob's  state is one of $\{\rho^{(k)}_{00}\}$. He
can perfectly discriminate them by a projective measurement since
they are orthogonal. The case when $m=1$ can be analyzed similarly.

Now we turn to show the necessity. Suppose that $\rho_1,\cdots,
\rho_N$ can be discriminated with certainty when Alice goes first.
Let us assume that Alice's nontrivial and orthogonality-keeping
measurement operator be $M_m$. Then for any $\ket{\psi_k}\in
Supp(\rho_k)$ and $\ket{\psi_l}\in Supp(\rho_l)$, we have
\begin{eqnarray}
\braket{\psi_k}{M_m^\dagger M_m\otimes I}{\psi_l}&=&0,\\
\ip{\psi_k}{\psi_l}&=&0,
\end{eqnarray}
where the second equation is due to the fact that $\rho_k$ and
$\rho_l$ are orthogonal. Let
\begin{equation}
M_m^\dagger M_m=\alpha \op{0}{0}+\beta\op{1}{1}
\end{equation}
be the spectral decomposition, where $\alpha>\beta\geq 0$ by the
non-triviality. Rewrite
\begin{eqnarray}
\ket{\psi_k}&=&\ket{0}\ket{\psi^{(k)}_0}+\ket{1}\ket{\psi^{(k)}_1},\\
\ket{\psi_l}&=&\ket{0}\ket{\psi^{(l)}_0}+\ket{1}\ket{\psi^{(l)}_1}.
\end{eqnarray}
Then we have
\begin{eqnarray}
\alpha\ip{\psi^{(k)}_0}{\psi^{(l)}_0}+\beta\ip{\psi^{(k)}_1}{\psi^{(l)}_1}&=&0,\\
\ip{\psi^{(k)}_0}{\psi^{(l)}_0}+\ip{\psi^{(k)}_1}{\psi^{(l)}_1}&=&0.
\end{eqnarray}
Since $\alpha\neq \beta$, the above system of equations has a unique
solution
$\ip{\psi^{(k)}_0}{\psi^{(l)}_0}=\ip{\psi^{(k)}_1}{\psi^{(l)}_1}=0$.
With that we complete the proof of the necessity. \hfill
$\blacksquare$\\

Intuitively speaking, the above theorem shows that a set of
$2\otimes n$ states are locally distinguishable when Alice goes
first if and only after Alice performs some suitable projective
measurement, the post-measurement states remain orthogonal.
Actually, we have proven the following
\begin{corollary}\label{2nconsequence}\upshape
If Alice goes first, then $\rho_1,\cdots, \rho_N$ can be perfectly
discriminated by LOCC if and only if there exists an orthogonal
basis $\{\ket{0},\ket{1}\}_A$ such that each $\rho_k$ has the form:
\begin{eqnarray}
\rho_k=\sum_{m,n=0}^1 \op{m}{n}\otimes \rho_{mn}^{(k)},
\end{eqnarray}
where $\rho_{mm}^{(1)},\cdots, \rho_{mm}^{(N)}$ are pairwise
orthogonal for each $m=0,1$.
\end{corollary}

The decomposition in the Eq. (\ref{nice-form}) can be analytically
determined.

\section{Local projective measurements and classical communication are sufficient for locally distinguishing multi-qubit states}

A simple but remarkable consequence of Theorem \ref{2otimesn} is
that local projective measurement and classical communication is
powerful enough to locally distinguish a set of multi-qubit
orthogonal quantum states. Due to its significance, we formally
state it as follows:
\begin{theorem}\label{manyqubits}\upshape
Local projective measurements and classical communications are
sufficient for deciding the local distinguishability of any set of
multi-qubit orthogonal quantum states.
\end{theorem}

The above theorem simplifies the local distinguishability of
multi-qubit considerably and makes it almost feasible to locally
distinguish a set of multi-qubit orthogonal states. A procedure can
be described as follows. Suppose $n$ qubits are held by
$Alice_1$,$\cdots$, $Alice_n$, respectively. Then any set of
$n$-qubit orthogonal states can be perfectly distinguishable by LOCC
if and only if they can be distinguishable by local projective
measurements and classical communication (LPMCC). Let $\sigma$ be a
permutation on $n$ qubits which is used to specify the order of the
projective measurement performed by each party. That is,
$Alice_{\sigma(1)}$ is the first person who performs a non-trivial
measurement, $Alice_{\sigma(2)}$ is the second one who performs a
conditional non-trivial measurement depending on
$Alice_{\sigma(2)}$'s outcome, $\cdots$, and $Alice_{\sigma(n)}$ is
the last one who performs conditional projective measurement
depending on the previous $n-1$ party's outcome. The projective
measurement performed by $Alice_{\sigma(1)}$ is given by an
orthogonal basis
$P_{\sigma(1)}(x_1)=\{\ket{\psi_{\sigma(1)}(x_1)}:x_1=0,1\}$, where
$x_1$ represents the outcome. According to the outcome of
$Alice_{\sigma_1}$, $Alice_{\sigma(2)}$ performs a projective
measurement
$P_{\sigma(2)}(x_1x_2)=\{\ket{\psi_{\sigma(2)}(x_1x_2)}:x_2=0,1\}$
with outcome $x_2$. Finally, depending on the previous outcomes
$x_1,\cdots, x_{n-1}$, $Alice_{\sigma(n)}$ performs a conditional
projective measurement $P_{\sigma(n)}(x_1x_2\cdots
x_n)=\{\ket{\psi_{\sigma(n)}(x_1x_2\cdots x_n)}: x_n=0,1\}$. The
above procedure induces a projective measurement on $n$ qubits
represented by an orthogonal product basis $\{\ket{\psi(x)}:x\in
\{0,1\}^n\}$, where
\begin{equation}
\ket{\psi(x)}=\ket{\psi_{\sigma(1)}(x_1)}\otimes \cdots\otimes
\ket{\psi_{\sigma(n)}(x_1x_2\cdots x_n)}.
\end{equation}
It is clear that $\rho_1,\cdots, \rho_N$ are perfectly
distinguishable by $Alice_1,\cdots, Alice_n$ if and only if they
lead to different (non-overlap) measurement outcomes. That is,
there exists a permutation $\sigma$, and a disjoint partition of
$\{0,1\}^n$, say $O_1,\cdots, O_N$ such that
\begin{equation}\label{key-eq}
\supp(\rho_k)\subseteq {\rm Span}\{\ket{\psi(x)}: x\in O_k\}.
\end{equation}
For each $\sigma$, the sequence of projective measurement, say,
$P_{\sigma(1)(x_1)},\cdots, P_{\sigma(n)}(x_1\cdots x_n)$ can be
analytically determined. Repeating the above process for all the
$n!$ permutations, we can completely determine the local
distinguishability of $\{\rho_1,\cdots,\rho_N\}$.

Eq. (\ref{key-eq}) implies a simple but highly nontrivial criterion
for local distinguishability of a set of multi-qbut states.
\begin{theorem}\label{schmidt-number-sum}\upshape
Let $S=\{\rho_1,\cdots, \rho_N\}$ be a collection  of orthogonal
states on $n$ qubits. Then $S$ is perfectly distinguishable by LOCC
only if the sum of the orthogonal Schmidt numbers of $\rho_k$ is not
more than the total dimension of state space, i.e.,
\begin{equation}\label{schmidt}
\sum_{k=1}^N \sch_{\perp}(\rho_k)\leq 2^n.
\end{equation}
\end{theorem}

A few remarks come as follows. Eq. (\ref{schmidt}) reflects the fact
that local projective measurements and classical communication are
sufficient for locally distinguishing multi-qubit quantum states. In
general, only projective measurements are not able to distinguish
quantum states locally. For such peculiar states the summations of
the orthogonal Schmidt numbers may exceed the total dimension of the
state space. An explicit instance of such peculiar states has been
found by Cohen very recently \cite{COH07}. We would also like to
point out that Eq. (\ref{schmidt}) is not sufficient when the number
of qubits under consideration are more than $2$: There do exist four
$2\otimes 2\otimes 2$ orthogonal product pure states that are
indistinguishable by LOCC \cite{BDM+99}.

To appreciate the power of Theorems \ref{manyqubits} and
\ref{schmidt-number-sum}, we shall present two examples concerning
with the local distinguishability of GHZ-type and W-type states,
which are respectively defined as
\begin{equation}\label{GHZ}
\alpha\ket{00\cdots 0}+\beta\ket{11\cdots 1}
\end{equation}
and
\begin{equation}\label{W}
\alpha_1\ket{00\cdots 1}+\alpha_2\ket{0\cdots
01}+\cdots+\alpha_n\ket{10\cdots 0},
\end{equation}
where each $\alpha_k$ is nonzero complex number and $n\geq 2$. It
has been shown that any $n$-qubit W-type state has orthogonal
Schmidt number $n$. By Theorem \ref{schmidt-number-sum}, there are
at most $2^n/n$ W-type states can be locally distinguishable. In
particular, any three $3$-qubit W-type states are locally
indistinguishable.

For GHZ-type states we shall show that any three $n$-qubit states
containing at least two GHZ-type states of the form Eq. (\ref{GHZ})
are locally indistinguishable ($n\geq 2$). The proof is by
mathematical induction. First, the base case when $n=2$ directly
follows from Theorem \ref{schmidt-number-sum} as the summation of
the orthogonal Schmidt numbers is at least $2+2+1=5>4$. Second,
suppose the result is valid for $n-1$, we will show the result is
also valid for $n$. By Theorem \ref{manyqubits}, we only need to
consider local projective measurements. Assume some party performs a
projective measurement $\{\ket{\psi},\ket{\psi^\perp}\}$. If
$\ket{\psi}\in \{\ket{0},\ket{1}\}$ (up to some phase factor), then
after the measurement two GHZ-type states will reduce to identical
states such as $\ket{0}^{\otimes n-1}$ or $\ket{1}^{\otimes n-1}$
thus cannot be further distinguished. If $\ket{\psi}\not \in
\{\ket{0},\ket{1}\}$, then at least for one measurement outcome ($0$
or $1$) the possible remaining states are three $n-1$-qubit states
containing at least two-GHZ type state. The proof is completed by
applying induction hypothesis. We notice the case of $n=3$ has been
solved by Ye {\it et al.} using a different but much more
complicated  method \cite{YJC+06}.

\section{Local distinguishability of $2\otimes 2$ states}
When only $2\otimes 2$ states are under consideration, Eq.
(\ref{schmidt}) is also sufficient for the local distinguishability.
\begin{theorem}\label{2qubits}\upshape
Let $S=\{\rho_1,\cdots,\rho_N\}$ be a set of $2\otimes 2$ orthogonal
states. Then $S$ is locally distinguishable if and only if
$\sum_{k=1}^N \sch_{\perp}(\rho_k)\leq 4$.
\end{theorem}

\textbf{Proof.} If each $\rho_k$ is a pure state, then the above
theorem is reduced to the one given by Walgate and Hardy
\cite{WH02}. Here we need to consider the case when some state may
be mixed.  We consider three cases according the number of states.
First we consider the case of $N=2$. By Theorem \ref{schmidt}, we
only need to consider the sufficiency. If both $\rho_1$ and $\rho_2$
are mixed states. Then we should have
$\sch_\perp(\rho_1)=Sch_\perp(\rho_2)=2$. That means there exists a
set of orthogonal product states $\{\ket{\Psi_k}:1\leq k\leq 4\}$
such that
\begin{eqnarray}
\supp(\rho_1)&=&{\rm Span}\{\ket{\Psi_1},\ket{\Psi_2}\},\\
\supp(\rho_3)&=&{\rm Span}\{\ket{\Psi_3},\ket{\Psi_4}\}.
\end{eqnarray}
Applying the result of Walgate and Hardy \cite{WH02}, we know that
Alice and Bob can  perfectly distinguish $\{\ket{\Psi_k}:1\leq k\leq
4\}$. If the outcome is $1$ or $2$, then the state is $\rho_1$;
otherwise is $\rho_2$. Using similar arguments, we can prove the
case when only one of $\rho_1$ and $\rho_2$ is mixed.

The case of three or four states is rather simple. Actually, we can
show that three $2\otimes 2$ orthogonal states are locally
distinguishable if and only if there are two product states, and
four $2\otimes 2$ orthogonal states are locally distinguishable if
and only if they are all product pure states. These results are
completely in accordance with Ref. \cite{WH02}\hfill $\blacksquare$

Theorem \ref{2qubits} indicates the local distinguishability of a
set of $2\otimes 2$ orthogonal states is completely characterized by
their orthogonal Schmidt numbers. More precisely, a set of states
$\{\rho_1,\cdots,\rho_N\}$ ($2\leq N\leq 4$) is locally
distinguishable the set orthogonal Schmidt numbers belongs to one of
the following case: $\{2,2\}$, $\{2,1,1\}$, $\{1,1,1,1\}$,
$\{2,1\}$, $\{1,1,1\}$, $\{1,1\}$.

Employing Theorem \ref{2qubits}, we can show there exists pairs of
orthogonal $2\otimes 2$ quantum states that are locally
indistinguishable. Let $\rho_1$ be a uniform mixture of $\ket{00}$
and $\ket{++}$, and let $\rho_2$ be a uniform mixture of $\ket{1-}$
and $\ket{-1}$. It is clear that both $\rho_1$ and $\rho_2$ are
separable. However, we have
$$\sch_\perp(\rho_1)=\sch_{\perp}(\rho_2)=3,$$
which immediately implies that $\rho_1$ and $\rho_2$ are locally
indistinguishable as the sum exceeds $4$. The indistinguishability
between $\rho_1$ and $\rho_2$ has already been proven in Ref.
\cite{BDF+99}, and was thoroughly studied in the scenario of quantum
data hiding \cite{DLT02}. But here we supply a rather different
approach which is of independent interest. Similarly, let
$\rho_3=\op{\psi}{\psi}$ such that $\ket{\psi}=\alpha
\ket{1-}+\beta\ket{-1}$ and $\alpha\beta\neq 0$, then $\rho_1$ and
$\rho_3$ are indistinguishable as
$\sch_\perp(\rho_1)+\sch_\perp(\rho_3)=5>4$.
\section{Locally indistinguishable subspaces spanned by $2\otimes 2\otimes 2$ unextendible product bases}

Now we begin to study three-qubit quantum systems. First we present
a formal definition of locally indistinguishable subspace.
\begin{definition}\label{IDS}\upshape
Let $S$ be a subspace of $\mathcal{H}=\otimes_{k=1}^m
\mathcal{H}_k$. $S$ is said to be locally indistinguishable if any
orthogonal basis of $S$ cannot be perfectly distinguishable by LOCC.
\end{definition}

The first instance of locally indistinguishable subspace was given
by Watrous \cite{WAT05}. For completeness, we give a short review
here. Let $\ket{\Phi}$ be a $d\otimes d$ maximally entangled state,
then it was proven that $\{\ket{\Phi}\}^\perp$ is locally
indistinguishable whenever $d>2$. Actually, what was shown in Ref.
\cite{WAT05} is that $\{\ket{\Phi}\}^\perp$ has no orthonormal basis
perfectly distinguishable by separable operations rather than LOCC.
With this subspace, Watrous constructed a class of quantum channels
which have sub-optimal environment-assisted capacity thus solved an
open problem suggested by Hayden and King \cite{HK05}. Clearly, the
minimal dimension of indistinguishable subspaces obtained by Watrous
is $3^2-1=8$. In Ref. \cite{DFXY07} we generalized this result to
arbitrary multipartite pure state $\ket{\Psi}$ with orthogonal
Schmidt number not less than $3$. Furthermore, we explicitly
constructed an indistinguishable bipartite subspaces with dimension
$3^2-2=7$. Our method also gave a $2\otimes 2 \otimes 2$
indistinguishable subspace with dimension $2^3-2=6$. How to further
reduce the dimension of indistinguishable subspaces remains a
difficult problem. In particular, we still don't know whether there
are locally indistinguishable subspaces with dimensions $3$, $4$,
$5$, respectively.

All the known indistinguishable subspaces up to now are not only
indistinguishable by LOCC, but also indistinguishable by separable
operations. So a question of interest naturally arises: Does  there
exist some indistinguishable subspace which is locally
indistinguishable but has  some orthonormal basis distinguishable by
separable operations? If such subspace does exist, we would expect
it has a smaller dimension.

In what follows we show that locally indistinguishable subspaces
with dimensions $3$, $4$, and $5$ do exist. Moreover, subspaces with
dimension $4$ can be constructed from $2\otimes 2\otimes 2$
orthogonal UPB and have bases distinguishable by separable
operations. We discuss the case of dimension $4$ here, and the other
two cases will be discussed in next section.

Note that any UPB for three-qubits should have four members, and can
be uniquely written into the following from (up to some local
unitary) \cite{BRA03}:
\begin{eqnarray}\label{3UPB}
\ket{S_1}&=&\ket{0}\otimes \ket{0}\otimes \ket{0},\nonumber \\
\ket{S_2}&=&\ket{1}\otimes \ket{B}\otimes \ket{C},\nonumber \\
\ket{S_3}&=&\ket{A}\otimes \ket{1}\otimes \ket{C^{\perp}},\nonumber \\
\ket{S_4}&=&\ket{A^{\perp}}\otimes \ket{B^{\perp}}\otimes \ket{1},
\end{eqnarray}
where $\ket{A}=\cos\theta_1\ket{0}+\sin\theta_1\ket{1}$,
$\ket{B}=\cos\theta_2\ket{0}+\sin\theta_2\ket{1}$,
$\ket{C}=\cos\theta_3\ket{0}+\sin\theta_3\ket{1}$, and
$\theta_1,\theta_2,\theta_3\in (0,\pi/2)$. Our main result is the
following
\begin{theorem}\label{main}\upshape
Let $S$ be the subspace spanned by the UPB defined in Eq.
(\ref{3UPB}). Then $S$ is locally indistinguishable, but has an
orthogonal basis distinguishable by separable operations.
\end{theorem}

{\bf Proof.} Applying the results in Ref. \cite{DMS+00} or Ref.
\cite{DFXY07}, we can easily see that the set of $\{S_k:1\leq k\leq
4\}$ is perfectly distinguishable by separable operations. That
completes the proof that $S$ has a basis distinguishable by
separable operations.

Now we show that any basis of $S$ is locally indistinguishable. It
is easy to see that any orthonormal basis $\{\ket{\Phi_k}:1\leq
k\leq 4\}$ of $S$ is uniquely determined by a $4\times 4$ unitary
matrix $U$ as follows:
\begin{equation}
\ket{\Phi_k}=u_{k1}\ket{S_1}+u_{k2}\ket{S_2}+u_{k3}\ket{S_3}+u_{k4}\ket{S_4},~~1\leq
k\leq 4.
\end{equation}
We shall consider five cases to complete the proof. By the symmetry,
we may assume that Alice is the one who goes first. The basic idea
is to show after Alice performs a projective measurements
$\{\ket{\psi},\ket{\psi^{\perp}}\}$, the post-measurement states are
indistinguishable. It is worth noting that we shall employ an
interesting property about $S$: There are only four product states
in $S$. A proof of this fact was given in Ref. \cite{BRA03}. One can
also prove it by a direct calculation.

Case 1. There are four product states in $\{\ket{\Phi_k}:1\leq k\leq
4\}$. They are just the UPB $\{S_k:1\leq k\leq 4\}$ , thus are
locally indistinguishable \cite{BDM+99}. A direct proof is as
follows. If $\ket{\psi}\not \in \{\ket{0},\ket{A}\}$, then the left
states are four nonorthogonal product states, which is
indistinguishable. Even if $\ket{\psi}\in \{\ket{0},\ket{A}\}$, the
left states are not orthogonal.

Case 2. There are three product states in $\{\ket{\Phi_k}:1\leq
k\leq 4\}$. This case cannot happen.

Case 3. There are two product states. Without loss of generality,
assume that $\{\ket{\Phi_k}:1\leq k\leq 4\}$ is of the following
form:
\begin{eqnarray}
\ket{S_1},~~~\ket{S_2},\\
u_1\ket{S_3}+u_2\ket{S_4},\\
u_2^*\ket{S_3}-u_1^*\ket{S_4}
\end{eqnarray}
where $|u_1|^2+|u_2|^2=1$ and $u_1u_2\neq 0$. If
$\ket{\psi}=\ket{0}$ then the left (unnormalized) states are
\begin{eqnarray}
\ket{00},\\
u_1\ip{0}{A}\ket{1C^\perp}+u_2\ip{0}{A^\perp}\ket{B^\perp
1},\\
u_2^*\ip{0}{A}\ket{1C^\perp}-u_1^*\ip{0}{A^\perp}\ket{B^\perp 1},
\end{eqnarray}
which contains two entangled states and one product state. It
follows from Ref. \cite{WH02} that they cannot be perfectly
distinguishable by LOCC. If $\ket{\psi}=\ket{A}$ then the left
states are not orthogonal to each other. If $\ket{\psi}\not\in
\{\ket{0},\ket{1},\ket{A},\ket{A^\perp}\}$, then the left states are
not orthogonal to each other.

Case 4. There is a unique  product state. Similar to Case 3. The
orthogonality was ruined.

Case 5. There is no product states. This is the most nontrivial case
we need to discuss. After Alice performs a projective measurement,
we have four left states. They should constitute an orthogonal
product basis. That immediately implies  $\{\ket{\Phi_k}\}$ should
be of the following form:
\begin{equation}
\ket{\Phi_k}=\alpha_k\ket{\psi}\otimes
\ket{ab_k}+\beta_k\ket{\psi^{\perp}}\otimes\ket{cd_k},~1\leq k\leq
4,
\end{equation}
where $\alpha_k$ and $\beta_k$ are nonzero complex numbers such that
$|\alpha_k|^2+|\beta_k|^2=1$, and $\{\ket{ab}_k\}$ and
$\{\ket{cd}_k\}$ are two orthogonal product bases. However, we shall
show that such a representation is not possible.

First we derive a relation between $\{\ket{ab_k}\}$ and
$\{\ket{cd_k}\}$. Consider the set of product states
$\{\ket{00},\ket{BC},\ket{1 C^\perp 1},\ket{B^\perp 1}\}$. Let
$\{\ket{\widetilde{00}}, \ket{\widetilde{BC}}, \ket{\widetilde{1
C^\perp}}, \ket{\widetilde{B^\perp 1}}\}$ be its reciprocal basis.
First we have
\begin{eqnarray}
\alpha_k\ket{ab_k}&=&u_{k1}\ip{\psi}{0}\ket{00}+u_{k2}\ip{\psi}{1}\ket{BC}\\
                  &+&u_{k3}\ip{\psi}{A}\ket{1C^\perp}+u_{k4}\ip{\psi}{A^\perp}\ket{B^\perp 1},\nonumber\\
\beta_k\ket{cd_k}&=&u_{k1}\ip{\psi^\perp}{0}\ket{00}+u_{k2}\ip{\psi^\perp}{1}\ket{BC}\\
                 &+&u_{k3}\ip{\psi^\perp}{A}\ket{1C^\perp}+u_{k4}\ip{\psi^\perp}{A^\perp}\ket{B^\perp 1}.\nonumber
\end{eqnarray}
From the equation about $\ket{ab_k}$ we have:
\begin{eqnarray}
u_{k1}&=&\frac{\alpha_k}{\ip{\psi}{0}}\frac{\ip{\widetilde{00}}{ab_k}}{\ip{\widetilde{00}}{00}},\\
u_{k2}&=&\frac{\alpha_k}{\ip{\psi}{1}}\frac{\ip{\widetilde{BC}}{ab_k}}{\ip{\widetilde{BC}}{BC}},\\
u_{k3}&=&\frac{\alpha_k}{\ip{\psi}{A}}\frac{\ip{\widetilde{1C^\perp}}{ab_k}}{\ip{\widetilde{1C^\perp}}{1C^\perp}},\\
u_{k4}&=&\frac{\alpha_k}{\ip{\psi}{A^\perp}}\frac{\ip{\widetilde{B^\perp
1}}{ab_k}}{\ip{\widetilde{B^\perp 1}}{B^\perp 1}}.
\end{eqnarray}
Substituting these equations into the equation about $\ket{cd_k}$,
we have
\begin{equation}\label{M-eq1}
M\ket{ab_k}=\frac{\beta_k}{\alpha_k}\ket{cd_k},~1\leq k\leq 4,
\end{equation}
where
\begin{eqnarray}\label{M-eq2}
M&=&r\frac{\op{00}{\widetilde{00}}}{\ip{\widetilde{00}}{00}}-\frac{1}{r^*}\frac{\op{BC}{\widetilde{BC}}}{\ip{\widetilde{BC}}{BC}}\\
&+&s\frac{\op{1C^\perp}{\widetilde{1C^\perp}}}{\ip{\widetilde{1C^\perp}}{1C^\perp}}-\frac{1}{s^*}\frac{\op{B^\perp
1}{\widetilde{B^\perp 1}}}{\ip{\widetilde{B^\perp 1}}{B^\perp
1}},\nonumber
\end{eqnarray}
and $r=\frac{\ip{\psi^\perp}{0}}{\ip{\psi}{0}}$ and
$s=\frac{\ip{\psi^\perp}{A}}{\ip{\psi}{A^\perp}}$. A key observation
here is that from Eq. (\ref{M-eq1}) we can obtain a very useful form
of $M$ as follows:
\begin{eqnarray}\label{M-eq3}
M\ket{00}&=&r\ket{00},\\
M\ket{BC}&=&-1/r^*\ket{BC},\\
M\ket{1C^\perp}&=&s\ket{1C^\perp},\\
M\ket{B^\perp 1}&=&-1/s^*\ket{B^\perp 1}.
\end{eqnarray}

Second we turn to show how to determine $\ket{ab_k}$ and
$\ket{cd_k}$. Let $P$ be the projector on $S$, then
\begin{equation}\label{projector}
P=\sum_{k=1}^4 \op{S_k}{S_k}=\sum_{k=1}^4 \op{\Phi_k}{\Phi_k}.
\end{equation}

For simplicity, denote $p=|\ip{\psi}{0}|^2$, $q=|\ip{\psi}{A}|^2$,
and
\begin{eqnarray}
\rho_1&=&p\op{00}{00}+(1-p)\op{BC}{BC},\\
\rho_2&=&q\op{1C^\perp}{1C^\perp}+(1-q)\op{B^\perp 1}{B^\perp 1}.
\end{eqnarray}
Then we have
\begin{equation}\label{bridge1}
\braket{\psi}{P}{\psi}=\sum_{k=1}^4|\alpha_k|^2\op{ab_k}{ab_k}=\rho_1+\rho_2,
\end{equation}
which implies that $\ket{ab_k}$ is the eigenvector of
$\braket{\psi}{P}{\psi}$ associated with the eigenvalue
$|\alpha_k|^2$. The problem left is to find the spectral
decomposition of $\braket{\psi}{P}{\psi}$. It is clear that
$\rho_1\perp \rho_2$. So the spectral decomposition of
$\braket{\psi}{P}{\psi}$ is just the summation of the spectral
decompositions of $\rho_1$ and $\rho_2$. Suppose the spectral
decompositions of $\rho_1$ and $\rho_2$ are given as follows:
\begin{eqnarray}\label{rho1rho2}
\rho_1&=&\lambda(p)\op{\Psi_1}{\Psi_1}+(1-\lambda(p))\op{\Psi_2}{\Psi_2},\nonumber\\
\rho_2&=&\lambda(q)\op{\Psi_3}{\Psi_3}+(1-\lambda(q))\op{\Psi_4}{\Psi_4},
\end{eqnarray}
where
\begin{equation}\label{eigen-val}
\lambda(p)=\frac{1+\sqrt{1-4p(1-p)(1-c_2^2c_3^2)}}{2}.
\end{equation}
If $\lambda(p)\neq \lambda(q)$ or $1-\lambda(q)$, then
$\rho_1+\rho_2$ will have four distinct eigenvalues, namely
$\lambda(p)$, $\lambda(q)$, $1-\lambda(p)$, $1-\lambda(q)$, and four
unique entangled eigenvectors $\{\ket{\Psi_k}\}$. That means
$\braket{\psi}{P}{\psi}$ cannot have product states as its
eigenvectors, which contradicts Eq. (\ref{bridge1}). So we should
have $p=q$ or $p=1-q$. Without loss of generality, let us assume
$p=q$ and simply write $\lambda(p)$ as $\lambda$. By Eqs.
(\ref{bridge1}) and (\ref{rho1rho2}), it should hold that
\begin{eqnarray}
\op{\Psi_1}{\Psi_1}+\op{\Psi_3}{\Psi_3}&=&\op{ab_1}{ab_1}+\op{ab_2}{ab_2},\\
\op{\Psi_2}{\Psi_2}+\op{\Psi_4}{\Psi_4}&=&\op{ab_3}{ab_3}+\op{ab_4}{ab_4}.
\end{eqnarray}
In other words, $\ket{ab_1}$ and $\ket{ab_2}$ are just the two
unique orthogonal product states in
$span\{\ket{\Psi_1},\ket{\Psi_3}\}$, and similarly, $\ket{ab_3}$ and
$\ket{ab_4}$ are the other two unique orthogonal product states in
$span\{\ket{\Psi_2},\ket{\Psi_4}\}$. So if we can determine
$\{\ket{\Psi_k}\}$ then we can also determine $\ket{ab_k}$. To be
specific, we only show how to determine $\ket{ab_1}$ (assume the
corresponding eigenvalue is $\lambda$).

Let $\ket{\Psi_1}=\mu\ket{00}+\ket{BC}$ (unnormalized) be the
eigenvector of $\rho_1$ associated with eigenvalue $\lambda$, where
$\mu$ is some nonzero complex number. Then from
$\rho_1\ket{\Psi_1}=\lambda \ket{\Psi_1}$ we have
\begin{eqnarray}
p(\mu+\ip{00}{BC})=\lambda\mu,\\
(1-p)(\mu \ip{BC}{00}+1)=\lambda,
\end{eqnarray}
from which we know that both $\lambda$ and $\mu$ are nonzero real
numbers. More precisely, $\mu$ should satisfy the following
equation:
\begin{equation}\label{mu-eq1}
\mu^2-\mu \frac{1-r^2}{r^2 c_2c_3}-\frac{1}{r^2}=0,
\end{equation}
where we $c_k$ and $s_k$ as abbreviations for  $\cos\theta_k$ and
$\sin\theta_k$, respectively. Similarly, $\ket{\Psi_3}$ should also
be of the form $\mu\ket{1C^\perp}+\ket{B^\perp 1}$, where $\mu$ is
the same as that in $\ket{\Psi_1}$. This is simply  due to the fact
that $\ip{1C^\perp}{B^\perp 1}=\ip{00}{BC}$.  So we can write
$\ket{ab_1}$ (unnormalized) as
\begin{equation}\label{ab-1}
\ket{ab_1}=x(\mu\ket{00}+\ket{BC})+(\mu\ket{1C^\perp}+\ket{B^\perp
1}),
\end{equation}
where $x$ is a complex number needed to be determined. Substituting
$\ket{B}$, $\ket{C}$, $\ket{B^\perp}$, and $\ket{C^\perp}$ into the
above equation we have:
\begin{eqnarray}\label{ab-2}
\ket{ab_1}&=&x(\mu+c_2c_3)\ket{00}+(xc_2s_3+s_2)\ket{01}\\
          &+&(xs_2c_3+\mu
s_3)\ket{10}+(xs_2s_3-\mu c_3-c_2)\ket{11}.\nonumber
\end{eqnarray}
To guarantee that  $\ket{ab_1}$ is a product state, $x$ should
satisty the following equation:
\begin{equation}\label{x-eq1}
x^2-x\frac{(1+\mu^2)c_3+2\mu c_2}{\mu s_2 s_3}-1=0,
\end{equation}
which immediately implies that $x$ is a real number. The procedure
for determining $\ket{cd_1}$ is almost the same. Employing the
spectral decomposition of $\braket{\psi^\perp}{P}{\psi^\perp}$, we
can see that $\ket{cd_1}$ is an eigenvector of
$\braket{\psi^\perp}{P}{\psi^\perp}$ associated with eigenvalue
$1-\lambda$. (Here we have employed the fact that
$|\alpha_1|^2+|\beta_1|^2=1$ and $|\alpha_1|^2=\lambda$.)
Furthermore, we can similarly show that $\ket{cd_1}$ should be of
the following form:
\begin{equation}\label{cd-1}
\ket{cd_1}=y(\mu'\ket{00}+\ket{BC})+(\mu'\ket{1C^\perp}+\ket{B^\perp
1}),
\end{equation}
where $y$ is chosen in such a way so that the right hand side of the
above equation is a product vector. That means $y$ and $\mu'$ should
satisfy
\begin{equation}\label{y-eq1}
y^2-y\frac{(1+{\mu'}^2)c_3+2\mu' c_2}{\mu' s_2 s_3}-1=0.
\end{equation}
Applying Eqs. (\ref{M-eq1}) and (\ref{M-eq3}) (assume
$\xi=\frac{\beta_1}{\alpha_1}$) to Eqs. (\ref{ab-1}) and
(\ref{cd-1}) we have
\begin{equation}\label{bridge2}
rx\mu=\xi y\mu',~ -x/r^*=\xi y,~s\mu=\xi\mu',~ -1/s^*=\xi,
\end{equation}
which follows that
\begin{equation}
\mu'=-|r|^2 \mu=-|s|^2\mu, ~(r/s)^*=x/y.
\end{equation}
Noticing that $x$, $y$, and $r$ are all real, we have $s=\pm r$.
Assume
\begin{eqnarray}
\ket{\psi}&=&\sqrt{p}\ket{0}+e^{i\alpha}\sqrt{1-p}\ket{1},\nonumber\\
\ket{\psi^\perp}&=&\sqrt{1-p}\ket{0}-e^{i\alpha}\sqrt{p}\ket{1},
\end{eqnarray}
where $0\leq \alpha<2\pi$. Then by a direct calculation we have
\begin{equation}
r=\sqrt{\frac{{1-p}}{{p}}},~s=\frac{\sqrt{1-p}c_1-\sqrt{p}s_1
e^{-i\alpha}}{\sqrt{p}c_1+\sqrt{1-p}s_1 e^{-i\alpha}}.
\end{equation}
$s$  also can be rewritten into the following form:
$$s=\frac{r-t_1e^{-i\alpha}}{1+rt_1e^{-i\alpha}}.$$
If $r=s$, then it follows that $r^2=-1$, a contradiction.  So
$r=-s$. We have
$$e^{i\alpha}=\frac{1-r^2}{2r}t_1.$$
$e^{i\alpha}=\pm 1$, then $(1-r^2)t_1=\pm 2r$. We also have $y=-x$
and $\mu'=-r^2\mu$. Substituting these two equations into Eq.
(\ref{y-eq1}) we have
\begin{equation}\label{x-eq2}
x^2-x\frac{(1+r^4{\mu}^2)c_3-2\mu r^2 c_2}{r^2 \mu s_2 s_3}-1=0.
\end{equation}
Comparing  Eqs. (\ref{x-eq1}) and (\ref{x-eq2}) we have
$$\frac{(1+r^4{\mu}^2)c_3-2\mu r^2 c_2}{r^2 \mu s_2 s_3}=\frac{(1+\mu^2)c_3+2\mu c_2}{\mu s_2 s_3},$$
or
\begin{equation}\label{mu-eq2}
\mu^2+ \mu \frac{4c_2}{c_3(1-r^2)}-\frac{1}{r^2}=0.
\end{equation}
By comparing Eqs. (\ref{mu-eq1}) and (\ref{mu-eq2}) we have
$$\frac{4c_2}{c_3(1-r^2)}=-\frac{1-r^2}{r^2 c_2c_3},$$
or
$$(\frac{1-r^2}{r})^2=-4c_2^2,$$
which is impossible for real number $r$. With that we complete the
proof. \hfill $\blacksquare$
\section{Unextendibility is not sufficient for local indistinguishability}
The notion of unextendible bases (UB) is a generalization of UPB
\cite{DFJY07}. Let $S$ be a set of linearly independent pure states
such that $S^\perp$ contains no product state, then we say $S$ is a
UB. Furthermore, if $S$ is a UB and any proper subset of $S$ cannot
be a UB, then we say $S$ is a genuinely UB (GUB). It has been shown
that the notion of UB is directly connected the local unambiguous
distinguishability. That is, a UB $S$ is distinguishable by
probabilistic LOCC if and only if  it is a GUB \cite{DFJY07}.

Since orthogonal UPB is a special case of GUB, a natural question is
whether any GUB can be used to construct locally indistinguishable
subspace. The answer is definitely no. Actually, the notion of UB is
introduced to characterize the distinguishability of quantum states
by probabilistic LOCC. So there is no direct relation between UB and
local (perfect) indistinguishability. Let us consider the following
four (unnormalized) states:
\begin{eqnarray}\label{UB1}
\ket{\Phi_1}&=&\ket{000}+e^{i\theta_1}\ket{111},\nonumber \\
\ket{\Phi_2}&=&\ket{001}+e^{i\theta_2}\ket{110},\nonumber\\
\ket{\Phi_3}&=&\ket{010}+e^{i\theta_3}\ket{101},\nonumber\\
\ket{\Phi_4}&=&\ket{011}+e^{i\theta_4}\ket{100},
\end{eqnarray}
where $0\leq \theta_1\leq \theta_2\leq \theta_3\leq \theta_4<2\pi$
and $\theta_4-\theta_3\neq \theta_2-\theta_1$. By a direct
calculation one can readily check that $\{\ket{\Phi_k}:1\leq k\leq
4\}$ is a UB. However, this set of states is clearly distinguishable
by LOCC. Each party only needs to perform a projective measurements
according to the standard basis $\{\ket{0},\ket{1}\}$. Another
interesting set of states which spans the orthogonal complement of
$\{\ket{\Phi_k}\}$ is as follows:
\begin{eqnarray}\label{UB2}
\ket{\Psi_1}&=&\ket{000}-e^{i\theta_1}\ket{111}, \nonumber\\
\ket{\Psi_2}&=&\ket{001}-e^{i\theta_2}\ket{110}, \nonumber\\
\ket{\Psi_3}&=&\ket{010}-e^{i\theta_3}\ket{101}, \nonumber\\
\ket{\Psi_4}&=&\ket{011}-e^{i\theta_4}\ket{100},
\end{eqnarray}
where $\{\theta_k\}$ is the same as the above equation. Then
$\{\ket{\Psi_k}\}$ is also an unextendible bases. Furthermore, let
$\rho_1$ and $\rho_2$ be two orthogonal mixed states with supports
$span\{\ket{\Phi_k}\}$ and $span\{\ket{\Psi_k}\}$, respectively.
Then it follows from Ref. \cite{CHE04} that $\rho_1$ and $\rho_2$
cannot be unambiguously distinguishable by LOCC. That is, we cannot
locally discriminate one of $\{\rho_1,\rho_2\}$ from the other with
a nonzero success probability without introducing error. Using the
terminology introduced in Ref. \cite{BW06}, we can say that none of
$\rho_1$ and $\rho_2$ is unambiguously identifiable by stochastic
local operations and classical communication (SLOCC). This result is
remarkable as it indicates that the local distinguishability of
mixed states are rather different from pure states, for which it has
been shown that any three linearly independent pure states are SLOCC
distinguishable \cite{BW06}. It would interesting to find similar
instance in bipartite scenario.

\section{Locally indistinguishable subspace with dimensions $3$}
Another worthwhile problem is whether there is a locally
indistinguishable subspace with dimension $3$. Note that any two
orthogonal pure states are locally distinguishable. Thus the
dimension of a locally indistinguishable subspace is at least $3$.
For $3\otimes 3$ quantum system it has been conjectured by King and
Matysiak that any subspace with dimension $3$ should contain a
locally distinguishable orthonormal basis \cite{KM05}. If this
conjecture were true, it would immediately imply that the dimension
of any $3\otimes 3$ locally indistinguishable subspace should at
least be $4$. However, for three-qubit system, it is possible to
construct a locally indistinguishable subspace with dimension $3$.
An explicit instance is as follows:
\begin{eqnarray}\label{3states}
\ket{\psi_1}&=&\ket{000},\nonumber\\
\ket{\psi_2}&=&\ket{100}-\ket{010},\nonumber\\
\ket{\psi_3}&=&\ket{100}+\ket{010}+\ket{001}.
\end{eqnarray}
Let $S_3=span\{\ket{\psi_1},\ket{\psi_2},\ket{\psi_3}\}$. We have
the following result.
\begin{theorem}\label{3LID}\upshape
$S_3$ is a locally indistinguishable subspace.
\end{theorem}

{\bf Proof.} Note that above states are not normalized, any basis of
the subspace spanned by the above three orthogonal states can be
expressed as $\ket{\alpha}=(\alpha_1,\alpha_2,\alpha_3)$,
$\ket{\beta}=(\beta_1,\beta_2,\beta_3)$ and
$\ket{\gamma}=(\gamma_1,\gamma_2,\gamma_3)$ under current basis. To
distinguish orthogonal states, one of Alice, Bob, and Charlie can
only perform a projective measurement to its own system and the
orthogonality between post-measurement states should be preserved.
We need to consider the following three cases corresponding to who
will perform the first measurement.

Case 1: Alice perform the measurement first. Suppose Alice's
measurement is given by the basis $\{\ket{\psi},\ket{\psi^\perp}\}$,
where $\ket{\phi}=a^*\ket{0}+b^*\ket{1}$ and $a\neq0$. After the
measurement, if outcome is $0$, the post-measurement state for
$\ket{\alpha}$ becomes
$(a\alpha_1+b\alpha_2+b\alpha_3)\ket{00}+a(\alpha_3-\alpha_2)\ket{10}+a\alpha_3\ket{01}$.
Notice that the state is eliminated only if $\ket{\alpha}=0$, thus
for the same reason, no state is eliminated after Alice's
measurement.

The question is now reduced to distinguish three $2\otimes2$
orthogonal states. The condition for distinguishability is at least
two states are product states. Suppose the two sates are
$\ket{\alpha}$ and $\ket{\beta}$, and notice $\ket{\alpha}$ becomes
a product state only when $\alpha_3-\alpha_2=0$ or $\alpha_3=0$, so
we also have $\beta_3-\beta_2=0$ or $\beta_3=0$.

case 1.1: $\alpha_3-\alpha_2=0$ and $\beta_3-\beta_2=0$. After
measurement, the two states become $\ket{0\alpha'}$ and
$\ket{0\beta'}$. The only state orthogonal to them is
$\ket{1\gamma'}$, that is $\ket{\gamma}$ also becomes a product
state after measurement, thus $\gamma_3=0$. But for orthogonality
between $\ket{\alpha}$, $\ket{\beta}$ and $\ket{\gamma}$, we require
$\gamma_1=0$ and $3\gamma_3=2\gamma_2\neq0$.

case 1.2: $\alpha_3=0$ and $\beta_3=0$. After measurement, the
states become $\ket{\alpha'0}$ and $\ket{\beta'0}$. So again
$\ket{\gamma}$ becomes a product state, thus require
$\gamma_3-\gamma_2=0$. But only when $\gamma_3\neq0$ and
$\gamma_1=\gamma_2=0$, could the three states be orthogonal to each
other.

case 1.3: $\alpha_3=0$ and $\beta_3-\beta_2=0$. After measurement,
the two states become $\ket{\alpha'0}$ and $\ket{0\beta'}$. As they
are orthogonal to each other, at leat one of $\ket{\alpha'}$ and
$\ket{\beta'}$ equals to $\ket{1}$. So after measurement,
$\ket{\gamma}$ should be orthogonal to either $\ket{10}$ or
$\ket{01}$, thus $\gamma_3=0$ or $\gamma_3-\gamma_2=0$ and we are
back to above two subcases.

case 2: Bob does measurement first. In this case the theorem stands
as we notice if we exchange the position of Alice and Bob, the three
states stay the same.

case 3: Charles does measurement first. After measurement,
$\ket{\alpha}$ becomes
$(a\alpha_1+b\alpha_3)\ket{00}+a(\alpha_3-\alpha_2)\ket{01}+a(\alpha_3+\alpha_2)\ket{10}$.
Notice $\ket{\alpha}$ is eliminated only when $\ket{\alpha}=0$, thus
two states must become product states after measurement. Suppose
they are $\ket{\alpha}$ and $\ket{\beta}$, then the following
conditions should be satisfied: $\alpha_3+\alpha_2=0$ or
$\alpha_3-\alpha_2=0$, and $\beta_3+\beta_2=0$ or
$\beta_3-\beta_2=0$. There are three subcases and similar discussion
in case 1 can be applied here.\hfill $\blacksquare$
\section{Locally indistinguishable subspace with dimensions $5$}
It is relatively easier to construct a locally indistinguishable
subspace with dimension $5$. Consider the subspace
$S_5=span\{\ket{\psi_k}:1\leq \ket{\psi_k}\leq 5\}$ where
\begin{eqnarray}\label{5states}
\ket{\psi_1}&=&\ket{000},\nonumber\\
\ket{\psi_2}&=&\ket{001}-\ket{100},\nonumber\\
\ket{\psi_3}&=&\ket{110}-\ket{011},\nonumber\\
\ket{\psi_4}&=&\ket{001}+\ket{010}+\ket{100},\nonumber\\
\ket{\psi_5}&=&\ket{110}+\ket{101}+\ket{011}.
\end{eqnarray}

\begin{theorem}\label{5dim}\upshape
$S_5$ is locally indistinguishable.
\end{theorem}
{\bf Proof.} The proof is rather straightforward. The key is to show
the following claim: There is a unique product state $\ket{\psi_1}$
in $S_5$. We postpone the proof of this claim and see how it implies
our result. Let $\{\ket{\phi_k}:1\leq k\leq 5\}$ be arbitrary
orthonormal basis for $S_5$. Then there is at least four product
states. So
$$\sum_{k=1}\sch_\perp(\ket{\phi_k})\geq 2\times 4+ 1=9>8.$$
By Theorem \ref{schmidt-number-sum}, we know that
$\{\ket{\phi_k}:1\leq k\leq 5\}$ cannot be distinguished by LOCC.

Now we prove the above claim. A general state
$\ket{\phi}=\alpha_k\ket{\psi_k}$ is expressed as follows:
\begin{eqnarray}
\ket{\phi}&=&\ket{0}(\alpha_1\ket{00}+(\alpha_2+\alpha_4)\ket{01}+\alpha_4\ket{10}+(\alpha_5-\alpha_3)\ket{11}),\nonumber\\
            &+&\ket{1}((\alpha_4-\alpha_2)\ket{00}+\alpha_5\ket{01}+(\alpha_3+\alpha_5)\ket{10})\nonumber,
\end{eqnarray}
which is a product vector if and only if
\begin{eqnarray}
\alpha_1(\alpha_5-\alpha_3)&=&\alpha_4(\alpha_2+\alpha_4),\nonumber\\
\alpha_5(\alpha_3+\alpha_5)&=&0,\nonumber\\
\lambda(\alpha_1,\alpha_2+\alpha_4,\alpha_4,\alpha_5-\alpha_3)&=&(\alpha_4-\alpha_2,\alpha_5,\alpha_3+\alpha_5,0).\nonumber
\end{eqnarray}
However, from the above equations we can easily verify that
$\alpha_5=\alpha_3=\alpha_4=\alpha_2=\lambda=0$. That is exactly
$\ket{\psi_1}$. \hfill $\blacksquare$

\section{Conclusion}
The most challenging open problem is to prove (or disprove) that any
orthogonal UPB spans a locally indistinguishable subspace. Another
worthwhile problem is to explore the applications of locally
indistinguishable subspaces. It has been shown by Watrous that
bipartite locally indistinguishable subspaces can be used to
construct quantum channels with sub-optimal environment-assisted
capacity \cite{WAT05}. However, for multipartite subspace little is
known.

This work was partly supported by the National Natural Science
Foundation of China (Grant Nos. 60702080, 60736011, and 60621062),
the FANEDD under Grant No. 200755, the Hi-Tech Research and
Development Program of China (863 project) (Grant No. 2006AA01Z102),
and the National Basic Research Program of China (Grant No.
2007CB807901).

\end{document}